\documentclass[twocolumn,showpacs,preprintnumbers]{revtex4}
\usepackage{graphicx}
\usepackage{dcolumn}
\usepackage{bm}

\begin{document}

\title{Correlation criteria for Bell type inequalities and entanglement
detection}
\author{Che-Ming Li$^{1,2}$}
\author{Li-Yi Hsu$^{3}$}
\author{Wei-Yang Lin$^{1}$}
\author{Yueh-Nan Chen$^{1,4}$}
\author{Der-San Chuu$^{1}$}
\author{Tobias Brandes$^{5}$}
\affiliation{$^{1}$Department of Electrophysics, National Chiao Tung University, Hsinchu
30050, Taiwan}
\affiliation{$^2$Physikalisches Institut, Universit\"{a}t Heidelberg, Philosophenweg 12,
D-69120 Heidelberg, Germany}
\affiliation{$^{3}$Department of Physics, Chung Yuan Christian University, Chung-li
32023, Taiwan}
\affiliation{$^{4}${National Center for Theoretical Sciences, National Cheng Kung
University, Tainan, Taiwan}}
\affiliation{$^{5}$Institut f\"{u}r Theoretische Physik, Technische Universit\"{a}t
Berlin, Hardenbergstr. 36 D-10623 Berlin, Germany}
\date{\today }

\begin{abstract}
We provide a novel criterion for identifying quantum correlation, which
allows us to find connections between Bell type inequalities, entanglement
detection, and correlation. We utilize the criterion to construct witness
operators that can detect genuine multi-qubit entanglement with fewer local
measurements. The connection between identifications of quantum correlation
and Mermin's inequality is discussed. Detection of genuine four-level
tripartite entanglement with two local measurement settings is shown in the
same manner. Further, through the criterion of quantum correlation, we
derive a new Bell inequality for arbitrary high-dimensional bipartite
systems, which requires fewer analyses of the measured outcomes.
\end{abstract}

\pacs{03.67.Mn,03.65.Ud}
\maketitle

\textit{Introduction.---} Bell type inequalities \cite
{bell,chsh,mermin,collins} and entanglement witnesses (EW) \cite
{witness,bouren,toth} lie at the heart of entanglement verification for
quantum information processing \cite{db}. Recently, the stabilizer formalism has been
utilized to derive Bell type inequalities \cite{scarani} and EW for multi-qubit systems \cite{toth}. It has been shown that
detections of genuine entanglement (GE) around several types of stabilizer
states require only two local measurement settings, and it also has been
found that the stabilizer witnesses are closely related to Mermin-type Bell
inequalities \cite{toth}. However, connections between multilevel Bell type
inequalities and EW are still not clear. Besides, there
still lakes a general way to detect a genuine multi-level multipartite
entanglement with fewer local measurements.

In this work, we present a new type of criterion for identifying quantum
correlation (QC), which is helpful for the investigation on the subjects
mentioned above. Firstly, EW for detecting genuine
multi-qubit entangled states are presented, including detection of
entanglement for states close to ones with nonlocal stabilizing operators,
e.g., the four-qubit state \cite{fourstate}. Connections between
Mermin-type Bell inequalities \cite{mermin} and criteria of QC are discussed. Secondly, we generalize the utility of
correlation criteria and propose the first EW for detecting GE around a four-level
tripartite Greenberger-Horne-Zeilinger (GHZ) state with two local measurement settings. Finally, through
identifications of QC, we give a new Bell type inequality
for arbitrary high-dimensional bipartite systems with fewer analyses of the
measured outcomes.

\textit{Criteria of QC and detection of GE for qubits.---} We first illustrate the main
notion of our strategy by providing a EW to detect GE
around a four-qubit GHZ state. According to our knowledge of the physical
state which is represented in the eigenbasis of the Pauli matrix $\sigma _{z}$: $\left| \text{
GHZ}\right\rangle =(\left| 0000\right\rangle _{z}+\left| 1111\right\rangle
_{z})/\sqrt{2}$, where $\left| kkkk\right\rangle _{z}\equiv \left|
k\right\rangle _{1,z}\otimes \left| k\right\rangle _{2,z}\otimes \left|
k\right\rangle _{3,z}\otimes \left| k\right\rangle _{4,z}$, we give four
sets of correlators to describe the QC between \emph{a
specific party} and \emph{others}: 
\begin{eqnarray}
&&C_{0,n}=\sum_{k=0}^{1}(-1)^{k}P(v_{n}=k,\text{v}=0), \\
&&C_{1,n}=\sum_{k=0}^{1}(-1)^{k+1}P(v_{n}=k,\text{v}=3),
\end{eqnarray}
where $\text{v}\equiv \sum_{i=1,i\neq n}^{4}v_{i},$ $v_{n}$ denotes
the outcome of a measurement performed on the $n^{\text{th}}$ particle for $
n=1,..,4$, and $P(v_{n}=k,\sum_{i=1,i\neq n}^{4}v_{i}=3m)$ stands for a
joint probability for obtaining $v_{n}=k$ and $v_{i}=m$ for $m=0,1$. If
results of measurements reveal that $C_{0,n}C_{1,n}>0$, we are convinced
that the outcomes of measurements performed on the $n^{\text{th}}$ particle
are correlated with the ones performed on the rest \cite{p1}. Further, with
a \emph{prior} information about probabilities for outcomes of measurements
of a GHZ state, $\mathcal{I}_{z\text{,GHZ}}$: $P(v_{nm}=0)+P(v_{nm}=2)=1$,
where $v_{nm}\equiv v_{n}+v_{m}$; $n,m=1,...,4$, and $n\neq m$, we construct
following correlators to identify correlations between \emph{a specific group
}, which is composed of the $n^{\text{th}}$ party and the $m^{\text{th}}$
one, and \emph{another}: 
\begin{eqnarray}
&&C_{0,nm}=\sum_{k=0}^{1}(-1)^{k}P(v_{nm}=2k,\text{v}^{\prime }=0), \\
&&C_{1,nm}=\sum_{k=0}^{1}(-1)^{k+1}P(v_{nm}=2k,\text{v}^{\prime }=2),
\end{eqnarray}
where $\text{v}^{\prime }\equiv \sum_{i=1,i\neq n\neq m}^{4}v_{i}$. It
is clear that $C_{0,nm}C_{1,nm}>0$ for a pure GHZ state, which indicates the
subsystem composed of the $n^{\text{th}}$ and the $m^{\text{th}}$ parts are
correlated with another \cite{p1}.

We consider the sets of correlators $C_{0,n}$, $C_{1,n}$, $C_{0,mn}$, and $
C_{1,mn}$ as identifications of a four-qubit GHZ state under the the local
measurement setting $\sigma_{z}^{\otimes 4}$ and take a combination of these
correlators: $C^{(z)}=
\sum_{n=1}^{4}(C_{0,n}+C_{1,n})+\sum_{m=2}^{4}(C_{0,1m}+C_{1,1m})$, as one
criterion of QC. Moreover, each correlator can be
characterized by a formulation of Hermitian operators. For instance, $C_{0,k}
$ can be characterized by the operator $\hat{C}_{0,k}=(\hat{0}_{k}-\hat{1}
_{k})\hat{0}_{i}\hat{0}_{j}\hat{0}_{l}$ for $k,i,j,l=1,...,4$ and $k\neq
i\neq j\neq l$, where $\hat{p}_{q}\equiv \left| p\right\rangle
_{qq}\left\langle p\right|$ for $p=0,1$ and $q=1,...,4$. Thus, we derive the
correlator operator: $\hat{C}^{(\text{I})}=8(\hat{0}\hat{0}\hat{0}\hat{0}+\hat{1}\hat{1}
\hat{1}\hat{1})-\openone$, from $C^{(z)}$.

After introducing the first kind criterion involved $\hat{C}^{(\text{I})}$,
let us progress to present the second one for QC. One can
acquire the prior information, $\mathcal{I}_{x,\text{GHZ}}$: $
\sum_{m=0}^{2}P(\sum_{n=1}^{4}v_{n}=2m)=1$, from the wave function of a
four-qubit GHZ state which is represented in the eigenbasis of the Pauli matrix $\sigma _{x}$
, i.e., $\left| \text{GHZ}\right\rangle =\sum_{m,n,i,j=0}^{1}\delta \lbrack
(m+n+i+j)\text{mod}\:2,0]\left| mnij\right\rangle _{x}/2\sqrt{2}$. Thus,
through $\mathcal{I}_{x,\text{GHZ}}$, we formulate four sets of criteria
which correspond to the following projection operators for identifying the
correlations between the $k^{\text{th}}$ party and others: $\hat{C}
_{0,k}^{(x)}=(\hat{0}_{k}-\hat{1}_{k})(\hat{0}_{i}\hat{0}_{j}\hat{0}_{l}+
\hat{0}_{i}\hat{1}_{j}\hat{1}_{l}+\hat{1}_{i}\hat{0}_{j}\hat{1}_{l}+\hat{0}
_{i}\hat{1}_{j}\hat{1}_{l}),\hat{C}_{1,k}^{(x)}=(\hat{1}_{k}-\hat{0}_{k})(
\hat{1}_{i}\hat{1}_{j}\hat{1}_{l}+\hat{1}_{i}\hat{0}_{j}\hat{0}_{l}+\hat{0}
_{i}\hat{1}_{j}\hat{0}_{l}+\hat{0}_{i}\hat{0}_{j}\hat{1}_{l})$. Since the
expectation values of operators, $\hat{C}_{0,k}^{(x)}$ and $\hat{C}
_{1,k}^{(x)}$, are all positive for a pure GHZ state, we ensure that there
are correlations between outcomes under the measurement setting $\sigma
_{x}^{\otimes 4}$. We combine the four sets of correlators and deduce the
Hermitian operators: $\hat{C}_{1}^{(\text{II})}=\sum_{k=1}^{4}\sum_{m=0}^{1}
\hat{H}^{\otimes 4}\hat{C}_{m,k}^{(x)}\hat{H}^{\otimes 4}=4X_{1}X_{2}X_{3}X_{4}$, where $\hat{H}$ is the Hadamard operator \cite{db} and $X_{m}=\sigma _{x}$ for the $m^{\text{th}}$ party. Similarly, as for identifications of QC between each party and others, we find that the following
operators work as well as $\hat{C}_{1}^{(\text{II})}$: $\hat{C}_{2}^{(\text{
II})}=4Y_{1}Y_{2}Y_{3}Y_{4}$ where $Y_{m}$ is the Pauli matrix $\sigma_{y}$; and the set of
operators which involve permutations of $XXYY$: $\hat{C}_{3}^{(\text{II})}=-4X_{1}X_{2}Y_{3}Y_{4}$,..., $\hat{C}_{8}^{(\text{II})}=-4Y_{1}Y_{2}X_{3}X_{4}$.

Then, we combine both kinds of criteria of QC as a identification of a four-qubit GHZ state and utilize the the witness operator: $\mathcal{W}_{\text{GHZ}}=\tau \openone-(c_{1}\hat{C}^{(\text{I}
)}+c_{2}\hat{C}^{(\text{II})}_{k})
$, for $k=1,...,8$, to identify a state $\rho $ as an \emph{genuinely}
entangled one which is close to a four-qubit GHZ state if it follows the
condition: $\text{Tr}[\mathcal{W}_{\text{GHZ}}\rho
]<0$. The values of $\tau$ and $c_{1,2}$ can be determined by the
condition of a multi-qubit witness \cite{toth}: $\mathcal{W}_{\text{GHZ}}-\gamma\mathcal{W}_{\text{GHZ}_{\text{p}}}\geq0$ where $\gamma$ is some positive constant and $\mathcal{W}
_{\text{GHZ}_{\text{p}}}$ is the projector-based witness \cite{bouren}, and
we have $\tau=7$, $c_{1}=c_{2}=1$, and $\gamma=8$. Moreover, when a state mixes with white noise, $\rho =p
\openone/16+(1-p)\left| \text{GHZ}\right\rangle \left\langle \text{GHZ}
\right| $, it is identified as an GE which is close to a
GHZ state when $
p<4/11(\approx 0.363636)$. Through our knowledge of the $N$-qubit GHZ state
and the same way presented above, one can formulate sets of correlators to
identify correlations between a group composed of $m$ parties and the rest $
(N-m)$ parts. From these criteria, we can construct the witness operator
that detects states around a $N$-qubit GHZ state \cite{further}.

Let us proceed to consider a scenario of entanglement detection which
involves only the second kind criteria. For the absence of the first kind
criterion to identify correlations between two groups, we find that the
operator, $\mathcal{W}_{2}^{(\text{II})}=\tau \openone-(\hat{C}_{k}^{(\text{
II})}+\hat{C}_{k^{\prime }}^{(\text{II})})$ for $k^{\prime }\neq k$, cannot
satisfy $\mathcal{W}_{2}^{(\text{II})}-\gamma \mathcal{W}_{\text{GHZ}_{\text{
p}}}\geq 0$. However, if we add more two terms to the operator, it will be
the case. For instance, the EW, $\mathcal{W}_{6}^{(\text{II})}=4.5
\openone-\sum_{k=1}^{6}\hat{C}_{k}^{(\text{II})}$, can be used to detect
GE, and it tolerates mixing with white noise with $p<1/3$.
When a EW contains all of the operators, i.e., $\mathcal{W}_{8}^{(\text{
II})}=4\openone-\hat{M}$ where $\hat{M}=\sum_{k=1}^{8}\hat{C}_{k}^{(\text{II}
)}$, it gains a noise tolerance up to $p<1/2$. It's noticeable that the
operator $\hat{M}$ is equivalent to the Bell operator in Mermin-type
Bell inequality. As for detections of GE,
Mermin-type Bell inequalities involve only the second kind criterion for
identifying QC. For many-qubit cases, the Bell operator in
Mermin's inequality also contains only the criteria for identifying the
correlations between a specific party and others.

Further, we can construct EW to detect states around stabilizing states through criteria of QC, including witnesses for cluster and graph states. If a state is described by stabilizing operators rather than the state vector, we also can derive criteria of QC from these locally measurable operators.  For example,  $Z_{3}X_{4}Z_{5}$ is one of the stabilizing operators of a five-qubit cluster state \cite{toth}, and from which we can construct the following operators to specify the QC between the $3^{\text{rd}}$, the $4^{\text{th}}$, and the $5^{\text{th}}$ qubits under the local measurements $Z$, $X$, and $Z$ respectively: $\hat{C}_{0,k}'=(\hat{0}_{k}-\hat{1}_{k})(\hat{0}_{i}\hat{0}_{j}+\hat{1}_{i}\hat{1}_{j}),\hat{C}_{1,k}'=(\hat{1}_{k}-\hat{0}_{k})(\hat{0}_{i}\hat{1}_{j}+\hat{1}_{i}\hat{0}_{j})$, for $i,j,k=3,4,5$, where $\hat{0}_{i(j,k)}$ and $\hat{1}_{i(j,k)}$ have been presented by the eigenstates of corresponding observables. Please note that $Z_{3}X_{4}Z_{5}= \hat{C}_{0,k}'+\hat{C}_{1,k}'$. For a pure five-qubit cluster state, the expectation values of $\hat{C}_{0,k}'$ and $\hat{C}_{1,k}'$ are both greater than zero \cite{further}, then we know there are correlations embedded in this subsystem.
By combining and utilizing these correlators which are derived from each stabilizing operator, we can achieve genuine entanglement detections \cite{further}.

To show that the proposed scenario can be applied to detect entangled states
around a specific state with nonlocal stabilizing operators, let us consider
how to construct a witness operator for states around a four-qubit state $
\Psi ^{(4)}$ \cite{fourstate}, which is a superposition of the tensor
product of two maximally entangled two-qubit states and a four-qubit GHZ
state, $ \left| \Psi ^{(4)}\right\rangle =\frac{1}{\sqrt{3}}(\left|
0011\right\rangle_{z} +\left| 1100\right\rangle_{z} -\frac{1}{2}(\left|
0110\right\rangle _{z}+\left| 1001\right\rangle_{z} +\left|
0101\right\rangle_{z} +\left| 1010\right\rangle_{z} ))$. We formulate eight
sets of criteria for identifying QC between a specific
party and others. The first type identifications include the following four
sets of correlators: $\hat{C}_{0,m}^{(z)}=\hat{0}\hat{0}\hat{1}\hat{1}-X_{m}(
\hat{0}\hat{0}\hat{1}\hat{1})X_{m}$, and $\hat{C}_{1,m}^{(z)}=\hat{1}\hat{1}
\hat{0}\hat{0}-X_{m}(\hat{1
}\hat{1}\hat{0}\hat{0})X_{m}$, for $m=1,...,4$. Then, the second type criteria are formulated as: $\hat{C}
_{0n,k}^{(z)}=(\hat{0}_{2n+1}\hat{1}_{2n+2}-X_{k}(\hat{0}_{2n+1} \hat{1}
_{2n+2})X_{k})(\hat{0}_{2n\oplus 3}\hat{1}_{2n\oplus 4}+\hat{1} _{2n\oplus 3}
\hat{0}_{2n\oplus 4})$, and $\hat{C}_{1n,k}^{(z)}=(\hat{1}_{2n+1}\hat{0}
_{2n+2}-X_{k}(\hat{1}_{2n+1} \hat{0}_{2n+2})X_{k})(\hat{0}_{2n\oplus 3}\hat{1
}_{2n\oplus 4}+\hat{1} _{2n\oplus 3}\hat{0}_{2n\oplus 4})$, where $
k=(2n+1),(2n+2)$ for $n=0,1$; and the symbol ''$\oplus $'' behaves as the
addition of modulo $4$ when $n=1$ and as an ordinary addition when $n=0$ is
met. For invariance of the wave function presented in the eigenbasis of $
\sigma _{x}$ ($\sigma _{y}$), in analogy, we can construct $8$ sets of
Hermitian operators, $(\hat{C}_{0,m}^{(x(y))},\hat{C} _{1,m}^{(x(y))})$ and $
(\hat{C}_{0n,k}^{(x(y))},\hat{C}_{1,nk}^{(x(y))})$, through the replacement
of the index $z$ in above Hermitian operators by the index $ x$ ($y$) and
constructing the operators in the eigenbasis of $\sigma _{x(y)} $. The expectation values of the above operators are all positive
for the state $\Psi ^{(4)}$.

Then, we give the following witness operator to detect GE
for states close to a $\Psi ^{(4)}$ state: $\mathcal{W}_{\Psi ^{(4)}}=\tau
\openone-(\hat{C}^{(x)}+\hat{C}^{(y)}+\hat{C} ^{(z)})$, where $\hat{C}^{(i)}=
\hat{U}_{i}^{\otimes 4}\sum_{l=0}^{1}\big(5\sum_{m=1}^{4}\hat{C}
_{l,m}^{(i)}+\sum_{n=0}^{1}\sum_{k=2n+1}^{2n+2}\hat{C}_{ln,k}^{(i)}\big)(
\hat{U}_{i}^{\dag })^{\otimes 4}$, for $i=x,y,z$, $\hat{U}_{x}=\hat{H}$, $
\hat{U}_{y}=\left( 
\begin{array}{cc}
-i & i \\ 
1 & 1
\end{array}
\right) /\sqrt{2}$, $\hat{U}_{z}=\openone$, and $\tau=36.5$ such that $\mathcal{W}_{\text{GHZ}}-30\mathcal{W}_{\text{GHZ}_{\text{p}}}>0$.  Moreover, it
tolerates mixing with white noise if $ p<15/88 (\approx 0.170455)$. With
only three local measurement settings, the above condition for the tolerance
of the noise is applicable to a real experiment \cite{bouren}.

\textit{Detection of GE for four-level tripartite system.---} In
order to show further utilities of the proposed scenario, we proceed to
provide a witness to detect GE close to a four-level tripartite
GHZ state: $\left| \text{GHZ}_{4\text{x}3}\right\rangle
=1/2\sum_{l=0}^{3}\left| l\right\rangle _{1,z}\otimes \left| l\right\rangle
_{2,z}\otimes \left| l\right\rangle _{3,z}$. First of all, by a knowledge of
the wave function represented in the eigenbasis: $\left| l\right\rangle
_{j,z}$ for $j=1,2,3$, we have $9$ sets of correlators for identifying
QC between the $m^{\text{th}}$ party and others, and we
derive the following operator from the $n^{\text{th}}$ set of correlators: $
\hat{C}_{mn}^{(z)}=\sum_{k=0}^{3}(\hat{k}-\hat{s}_{kn})_{m}\hat{k}_{p}\hat{k}
_{q}$, for $n=1,...,9$; $m,p,q=1,2,3$, and $m\neq p\neq q$; where $\hat{s}
_{kn}=\hat{0},...,\hat{3}$; $\hat{k}\neq \hat{s}_{kn}$ and $\hat{s}_{kn}\neq 
\hat{s}_{k^{\prime }n}$ for $k\neq k^{\prime }$; and $\hat{C}_{mn}^{(z)}\neq 
\hat{C}_{mn^{\prime }}^{(z)}$ for $n\neq n^{\prime }$.

Secondly, from our knowledge to an alternative representation of a $\text{GHZ}_{4\text{x}3}$ state, $\left| \text{GHZ}_{4\text{x}3}\right\rangle=1/4
\sum^{3}_{k,l,r=0}\delta[(k+l+r)\text{mod}\:4,0]\left|
k\right\rangle_{1,f}\otimes\left| l\right\rangle_{2,f}\otimes\left|
r\right\rangle_{3,f}$, where $\left|
g\right\rangle_{j,f}=1/2\sum_{h=0}^{3}e^{-i2hg\pi/4}\left|
h\right\rangle_{j,z}$, we can deduce the following operator from the $n^{
\text{th}}$ criteria of 9 sets correlators to identify QC between the $m^{\text{th}}$ party and others: $\hat{C}
^{(f)}_{mn}=\sum^{3}_{k=0}(\hat{F}^{\dag})^{\otimes 3}(\hat{k}-\hat{s}
_{kn})_{m}\hat{V}_{klr}\hat{F}^{\otimes 3}$, where $\hat{F}
=1/2\sum_{h,g=0}^{3}e^{i2hg\pi/4}\left| h\rangle\langle g\right|$, $\hat{V}
_{klr}=\sum^{3}_{l,r=0}\delta[(k+l+r)\text{mod}\:4,0]\hat{l}_{p}\hat{r}_{q}$
, and definitions of $\hat{k}$, $\hat{s}_{kn}$, $m$, $p$, $q$, and $n$ are same as the
ones mentioned for $\hat{C}^{(z)}_{mn}$.

With the derived correlators, we provide the following EW
to detect genuine four-level tripartite entanglement for states close to
a $\text{GHZ}_{4\text{x}3}$ state \cite{p2}: $\mathcal{W}_{\text{GHZ}_{4
\text{x}3}}=34.13\openone-\sum^{3}_{m=1}\sum^{9}_{n=1}(\hat{C}^{(z)}_{mn}+
\hat{C}^{(f)}_{mn})$. Furthermore, when a state mixes with white noise, the
EW, $\mathcal{W}_{\text{GHZ}_{4\text{x}3}}$, detects GE
if $p<0.368$. Thus, two local measurement settings are sufficient to detect
genuine four-level tripartite entanglement around a $\text{GHZ}_{4\text{x}3}$ state.

\textit{Bell type inequality for arbitrary high-dimensional bipartite
systems.---} Our scenario for deriving Bell type inequality starts with
specifications of the criteria for QC. Then, we proceed to
verify that any local theory cannot reproduce the correlations embedded in a
entangled state. This approach is novel and opposite to the one which has
been presented \cite{collins}.

First, to specify the QC embedded in the maximally
entangled state of two $d$-dimensional parts, $\left| \psi _{d}\right\rangle
=1/\sqrt{d} \sum_{n=0}^{d-1}\left|
n\right\rangle_{1,z}\otimes\left|n\right\rangle_{2,z}$, we represent the
wave function in the following eigenbasis: $\left|l\right\rangle_{k,j}=1/
\sqrt{d}\sum^{d-1}_{m=0}e^{i2\pi m(l+n_{k}^{(j)})/d}\left|
m\right\rangle_{k,z}$, where $ n_{1}^{(1)}=0$, $n_{2}^{(1)}=1/4$, $
n_{1}^{(2)}=1/2$, and $n_{2}^{(2)}=-1/4$ correspond to four different local
measurements. From our knowledge of the four different representations of
the state $\psi_{d}$, we give four sets of correlators of QC: 
\begin{eqnarray}
&&C_{m}^{(12)}=P(v_{1}^{(1)}=(-m)\,\text{mod}\:d,v_{2}^{(2)}=m)  \nonumber \\
&&\quad\quad\quad-P(v_{1}^{(1)}=(1-m)\text{mod}\:d,v_{2}^{(2)}=m), \\
&&C_{m}^{(21)}=P(v_{1}^{(2)}=(d-m-1)\text{mod}\:d,v_{2}^{(1)}=m)  \nonumber
\\
&&\quad\quad\quad-P(v_{1}^{(2)}=(-m)\text{mod}\:d,v_{2}^{(1)}=m), \\
&&C_{m}^{(qq)}=P(v_{1}^{(q)}=(-m)\,\text{mod}\:d,v_{2}^{(q)}=m)  \nonumber \\
&&\quad\quad\quad-P(v_{1}^{(q)}=(d-m-1)\text{mod}\:d,v_{2}^{(q)}=m),
\end{eqnarray}
for $m=0,1,...,d-1$ and $q=1,2$. The superscripts, $(ij)$, $(i)$, and $(j)$,
indicate that the local measurements $V_{1}^{(i)}$ and $V_{2}^{(j)}$ have
been selected by the first party and the second one respectively. Thus, we
take the summation of all $C_{m}^{(ij)}$ 's, 
\begin{equation}
C_{d}=C^{(11)}+C^{(12)}+C^{(21)}+C^{(22)},
\end{equation}
where $C^{(ij)}=\sum_{m=0}^{d-1}C_{m}^{(ij)}$, as an identification of the
state $\psi_{d}$.

For a pure state $\psi _{d}$, the correlator $C_{m}^{(ij)}$ can be evaluated
analytically and are given by $C_{m}^{(ij)}=(\csc ^{2}(\pi /4d)-\csc
^{2}(3\pi /4d))/2d^{3}$, where $\csc(h)$ is the cosecant of $h$. Since $
C_{m}^{(ij)}>0$ for all $m$'s with any finite value of $d$, we ensure that
there are correlations between outcomes of measurements performed on the
state $\psi _{d}$ under four different local measurement settings.
Furthermore, we can evaluate the summation of all $C_{m}^{(ij)}$'s, and then
we have $ C_{d,\psi _{d}}=2(\csc ^{2}(\pi /4d)-\csc ^{2}(3\pi /4d))/d^{2}$.
One can find that $C_{d,\psi _{d}}$ is an increasing function of $d$. For
instance, if $d=3$, one has $C_{3,\psi _{3}}\simeq 2.87293$. In the limit
large $d$, we obtain, $\lim_{d\rightarrow \infty }C_{d,\psi _{d}}=(16/3\pi
)^{2}\simeq 2.88202$.

We proceed to consider the maximum value of $C_{d}$ for local hidden
variable theories. The following derivation is based on deterministic local
models which are specified by \emph{fixing} the outcome of all measurements.
This consideration is general since any probabilistic model can be converted
into a deterministic one \cite{percival}. Substituting a \emph{fixed} set, $(
\tilde{v}_{1}^{(1)},\tilde{v}_{2}^{(1)},\tilde{v}_{1}^{(2)},\tilde{v}
_{2}^{(2)})$, into $C^{(ij)}$, $C_{d}$ turns into $C_{d,\text{LHV}}=\delta
\lbrack (\tilde{v}_{1}^{(1)}+\tilde{v}_{2}^{(1)})\text{mod} \:d,0]-\delta
\lbrack -(\tilde{v}_{1}^{(1)}+\tilde{v}_{2}^{(1)})\text{mod}\:d,1]+\delta
\lbrack (\tilde{v}_{1}^{(1)}+\tilde{v}_{2}^{(2)})\text{mod}\:d,0]-\delta
\lbrack (\tilde{v}_{1}^{(1)}+\tilde{v}_{2}^{(2)})\text{mod}\:d,1]+\delta
\lbrack (\tilde{v}_{1}^{(2)}+\tilde{v}_{2}^{(2)})\text{mod}\:d,0]-\delta
\lbrack -(\tilde{v}_{1}^{(2)}+\tilde{v}_{2}^{(2)})\text{mod}\:d,1]+\delta
\lbrack -(\tilde{v}_{1}^{(2)}+\tilde{v}_{2}^{(1)})\text{mod}\:d,1]-\delta
\lbrack (\tilde{v}_{1}^{(2)}+\tilde{v}_{2}^{(1)})\text{mod}\:d,0]$, where $
\delta \lbrack x,y]$ represent the Kronecker delta symbol. There are three
non-vanishing terms at most among the four positive delta functions, and
there exist four cases for it, for example, one is that if $\delta \lbrack (
\tilde{v}_{1}^{(1)}+\tilde{v}_{2}^{(1)})\text{mod} \:d,0]=\delta \lbrack (
\tilde{v}_{1}^{(1)}+\tilde{v}_{2}^{(2)})\text{mod} \:d,0]=\delta \lbrack (
\tilde{v}_{1}^{(2)}+\tilde{v}_{2}^{(2)})\text{mod} \:d,0]=1$ is assigned, we
obtain $\tilde{v}_{2}^{(1)}=\tilde{v}_{2}^{(2)}$ and then deduce that $
\delta \lbrack -(\tilde{v}_{1}^{(2)}+\tilde{v}_{2}^{(1)})\text{mod}\:d,1]=0$
. We also know that there must exist one non-vanishing negative delta
function and three vanishing negative ones in the $C_{d,\text{LHV}}$ under
the same condition. In the example, the case is $\delta \lbrack (\tilde{v}
_{1}^{(2)}+\tilde{v}_{2}^{(1)})\text{mod}\:d,0]=1$. With these facts, we
conclude that $C_{d,\text{LHV}}\leq 2$. One can check other three cases for
the four positive delta functions, and then they always result in the same
bound. Thus, we realize that $C_{d,\psi _{d}}>C_{d,\text{LHV} }$ and the
QC are stronger than the ones predicted by the local
hidden variable theories.

A surprising feature of the new inequality is that the total number of joint
probabilities required by each of the presented correlation functions $
C^{(ij)}$ is only $2d$, which is much smaller than that in Ref. \cite{fu},
which is about $O( d^{2})$. It implies that the proposed correlation
functions only contain the dominant terms to identify correlations. Besides,
the proposed scenario is robust against to noise. For instance, if the
system is under the condition that $p<0.30604$, the QC can
be maintained for the limit of large $d$.

Furthermore, although we haven't known yet wether $C_{d}$ can be utilized to
construct EW for detecting arbitrary high-dimensional entanglements
around a state $\psi_{d}$, as regards the cases have been analyzed, they
work for entanglement detection. Take the case for $d=4$ as an example, the
EW, $\mathcal{W}_{\psi_{4}}=2.05\openone-\hat{C}_{4}$, where $\hat{C}
_{4}$ is the operator which involves $C_{4}$, can be used to detect
entanglement for states around $\psi_{4}$, and it tolerates mixing with
white noise if $ p<0.2881$ \cite{further}.

\textit{Conclusion.---} We have provided a novel and syncretic approach to
derive a new Bell type inequality for arbitrary high-dimensional bipartite
systems and to construct EW to detect GE around several types of entangled qubits with only a small
effort for local measurements. The connection between Mermin-type Bell
inequalities and the criteria of QC is discussed. We also
show its utility to detect GE around a four-level tripartite GHZ state with two local measurement settings, which help
investigation on detections of genuine multilevel and multipartite
entanglement in an efficient way.

We are indebted to J.-W. Pan and Z.-B. Chen for fruitful discussions and
comments. This work is supported partially by the National Science Council,
Taiwan under the grant numbers NSC 94-2112-M-009-019, NSC 94-2120-M-009-002,
and NSC-94-2112-M-033-006. The author LYH is also partially supported by
National Center for Theoretical Sciences.


\begin{references}
\bibitem{p1}If the $n^{\text{th}}$ party is uncorrelated with the rest, we recast $P(v_{n}=k,\text{v}=3m)$ as $P(v_{n}=k)P(\text{v}=3m)$, where $P(v_{n}=k)$ and $P(\text{v}=3m)$ denote the probabilities for obtaining results $v_{n}=k$ and $\text{v}=3m$ respectively. Then, we have $C_{0,n}=(P(v_{n}=0)-P(v_{n}=1))P(\text{v}=0)$ and $C_{1,n}=(P(v_{n}=1)-P(v_{n}=0))P(\text{v}=3)$.  By the law of conservation of probability, it turns out that $C_{0,n}C_{1,n}\leq 0$. Similarly, if the subsystem composed of the $n^{\text{th}}$ party and the $m^{\text{th}}$ one is uncorrelated with another, the measured outcomes satisfy $C_{0,nm}C_{1,nm}\leq 0$.
\bibitem{p2} To prove that $\mathcal{W}_{\text{GHZ}_{4\text{x}3}}$ is an EW to detect GE, firstly, we have to know that $\mathcal{W}_{\text{ GHZ}_{4\text{x}3},\text{p}}=\openone/4-\left| \text{GHZ}_{4\text{x}3}\right\rangle
\left\langle \text{GHZ}_{4\text{x}3}\right| $ is a EW which detects GE. It can be shown by the methods presented in Ref. \cite{bouren}. Secondly, we show that $\mathcal{W}_{\text{GHZ}_{4\text{x}3}}$ satisfies the following criterion to a EW to detect GE: $\mathcal{W}_{\text{GHZ}_{4\text{x}3}}-\gamma\mathcal{W}_{\text{GHZ}_{4\text{x}3},\text{p}}\geq0$. Our result shows that $\gamma=28.5$.
\end{references}
\end{document}